\documentclass[]{tMOP2e}

\newcommand{\calG}{\mathcal{G}}

\newcommand{\calH}{\mathcal{H}}
\newcommand{\calF}{\mathcal{F}}
\newcommand{\calP}{\mathcal{P}}
\newcommand{\calGU}{\mathcal{G}_{U}} 
\newcommand{\bfF}{\mathbf{F}}
\newcommand{\bfsig}{\mbox{\boldmath{$\sigma$}}}
\newcommand{\bfhatz}{\hat{\mathbf{z}}} 
\newcommand{\bfhatx}{\hat{\mathbf{x}}}
\newcommand{\bfhaty}{\hat{\mathbf{y}}}
\newcommand{\xbar}{\overline{x}}
\newcommand{\kbar}{\overline{\kappa}}

\begin{document}

\doi{10.1080/0950034YYxxxxxxxx}
 \issn{1362-3044}
\issnp{0950-0340} \jvol{00} \jnum{00} \jyear{2010} \jmonth{10 January}

\markboth{Ran Li and Frank Gaitan}{Journal of Modern Optics}

\title{Robust High-Fidelity Universal Set of Quantum Gates Through 
Non-Adiabatic Rapid Passage}

\author{Ran Li$^{a}$ and Frank Gaitan$^{b}$$^{\ast}$
\thanks{$^\ast$Corresponding author. Email: fgaitan@lps.umd.edu\vspace{6pt}}\\ 
\vspace{6pt}  $^{a}${\em{Department of Physics, Kent State University, Stark 
Campus, North Canton, OH 44720}};\\$^{b}${\em{Laboratory for Physical Sciences,
8050 Greenmead Dr, College Park, MD 20740}}\\
\vspace{6pt}\received{} }

\maketitle

\begin{abstract}
We show how a robust high-fidelity universal set of quantum gates can be 
implemented using a single form of non-adiabatic rapid passage whose parameters
are optimized to maximize gate fidelity and reward gate robustness. Each gate 
in the universal set is found to operate with a fidelity $\mathcal{F}$ in the 
range $0.99988 < \mathcal{F} < 0.99999$, and to require control parameters
with no more than $14$-bit ($1$ part in $10^{4}$) precision. Such precision is
within reach of commercially available arbitrary waveform generators, so that
an experimental study of this approach to high-fidelity universal quantum 
control appears feasible.\bigskip

\bigskip
\end{abstract}

\section{Introduction}
\label{sec1}

Through the accuracy threshold theorem \cite{ft1,ft2,ft3,ft4,ft5,ft6,ft7,ft8},
it is now known that, under appropriate conditions, an arbitrary quantum 
computation can be done with arbitrarily small error probability, even in the 
presence of noise and imperfect quantum gates. The theorem requires the
protection of computational data through the use of a suitable quantum error 
correcting code, and the use of fault-tolerant procedures to control the spread 
of errors 
during computation, measurement, and error correction. It also requires 
the availability of a sufficiently reliable universal set of unencoded quantum 
gates, where sufficiently reliable means each gate has an error probability 
$P_{e}$ that is smaller than an accuracy threshold $P_{a}$. The actual value 
of this threshold is model dependent, though for many, $P_{a}\sim 10^{-4}$ 
has become a rough-and-ready estimate, with gates anticipated to be 
approaching the accuracies needed for fault-tolerant quantum computing 
when $P_{e} < 10^{-4}$. The accuracy 
threshold theorem quantifies the accuracy required of a quantum gate if it is 
to be used in fault-tolerant quantum computing. However, this is not the only 
requirement a quantum gate must satisfy. Gate performance must also be 
robust against small variation of the parameters that specify the control field 
driving the quantum gate. One of the central challenges facing the field of 
quantum computing is determining how to implement a robust universal set 
of unencoded quantum gates for which all gate error probabilities satisfy 
$P_{e} < 10^{-4}$.

In previous work we have shown how controllable quantum interference effects
arising during a form of non-adiabatic rapid passage known as twisted rapid 
passage (TRP) \cite{fg1,zwan2,fg3} can be used to implement a non-adiabatic 
high-fidelity universal set of quantum gates \cite{lg1,lhg2,lhg1}. In 
Ref.~\cite{lg1}, all TRP-generated gates were implemented with error 
probabilities satisfying $P_{e}<10^{-4}$. Although this level of gate 
performance achieves the anticipated accuracies needed for fault-tolerant 
quantum computing, it was found that the performance of the one-qubit gates
in the TRP-generated universal set was not sufficiently robust. 
Specifically, to achieve such high-fidelity performance, the TRP sweep 
parameters had to be controlled to $1$ part in $10^{5}$ (viz.~$17$-bit) 
precision. This degree of precision is beyond the capabilities of commercially 
available arbitrary waveform generators which can only achieve $1$ part in 
$10^{4}$ ($14$-bit) precision \cite{tek}. For TRP to form the basis for 
high-fidelity universal quantum computation, a way must be found to 
enhance the robustness of  the TRP gates.

In this paper we present a general approach for enhancing the robustness 
of an arbitrary quantum gate, and apply this approach to the one-qubit gates 
belonging to the universal set produced using TRP. We show that the 
resulting gates only require TRP sweep parameters to be controllable to
 $14$-bit precision to operate with error probabilities (fidelities) in the 
range $6.27\times 10^{-5}<P_{e}< 4.62\times 10^{-4}$ ($0.99988 
< \mathcal{F} < 0.99999$). The \textit{central result\/} of this paper is 
that our robustness-enhancing procedure has yielded a universal set of 
quantum gates that operate with very high fidelity, and yet are sufficiently 
robustness to be within reach of commercially available arbitrary waveform 
generators \cite{tek}. As we shall see, three (two) of the five TRP gates have 
error probabilities satisfying $P_{e}<1.12\times 10^{-4}$ ($2.13\times10^{-4}
<P_{e} < 4.62\times 10^{-4}$), and so operate at (nearly at) the anticipated 
accuracy needed for fault-tolerant quantum computing. 

The structure of this paper is as follows. In Section~\ref{sec2} we briefly 
review: (i)~the necessary background on TRP; (ii)~the universal set of 
quantum gates we implement using TRP; and (iii)~the numerical simulations 
used to determine gate performance. Section~\ref{sec3} then describes 
the optimization procedure used to maximize gate fidelity while 
simultaneously rewarding robustness. The results of this optimization 
procedure are presented in Section~\ref{sec4} for each of the gates in 
the TRP-generated universal set. Finally, we summarize our results and 
make closing remarks in Section~\ref{sec5}.

\section{Background}
\label{sec2}

\noindent In an effort to make this paper more self-contained, this Section
briefly reviews needed background material on TRP. For a more detailed 
presentation, the reader is directed to Refs.~\cite{fg1,lg1,lhg1}.

\subsection{TRP and Controllable Quantum Interference} 
\label{sec2.1} 

\noindent To introduce TRP \cite{fg1,lhg1}, we consider a single-qubit 
interacting with an external control-field $\bfF (t)$ via the Zeeman
interaction $H_{z}(t) = -\bfsig\cdot\bfF (t)$, where $\sigma_{i}$ are 
the Pauli matrices ($i=x,y,z$). TRP is a generalization of adiabatic rapid 
passage (ARP) \cite{abra}. In ARP, the control-field $\bfF (t)$ is slowly 
inverted over a time $T_{0}$ such that $\bfF (t) =at\,\bfhatz + b\,\bfhatx$. 
In TRP, however, the control-field is allowed to twist in the $x$-$y$ plane 
with time-varying azimuthal angle $\phi (t)$, while simultaneously 
undergoing inversion along the $z$-axis:
$\bfF (t) = at\,\bfhatz + b\,\cos\phi (t)\,\bfhatx +b\,\sin\phi (t)\,\bfhaty $.
Here $-T_{0}/2\leq t\leq T_{0}/2$, and throughout this paper, we consider 
TRP with \textit{non-adiabatic\/} inversion. As shown in Ref.~\cite{lhg1}, 
the qubit undergoes resonance when 
\begin{equation} 
at -\frac{\hbar}{2}\frac{d\phi}{dt} = 0 . 
\label{rescon}
\end{equation} 
For polynomial twist, the twist profile $\phi (t)$ takes the form 
\begin{equation} 
\phi_{n}(t) = \frac{2}{n}Bt^{n} . 
\label{polytwist} 
\end{equation} 
In this case, Eq.~(\ref{rescon}) has $n-1$ roots, though only real-valued roots 
correspond to resonance. Ref.~\cite{fg1} showed that for $n\geq 3$, the 
qubit undergoes resonance multiple times during a \textit{single\/} TRP sweep: 
(i)~for all $n\geq 3$, when $B>0$; and (ii)~for odd $n\geq 3$, when $B<0$. 
For the remainder of this paper we restrict ourselves to $B>0$, and to 
\textit{quartic\/} twist for which $n=4$ in Eq.~(\ref{polytwist}). During 
quartic twist, the qubit passes through resonance at times $t=0,\pm
\sqrt{a/\hbar B}$ \cite{fg1}. It is thus possible to alter the time separating 
the resonances by varying the TRP sweep parameters $B$ and $a$.

Ref.~\cite{fg1} showed that these multiple resonances have a strong 
influence on the qubit transition probability, allowing transitions to be
strongly enhanced or suppressed through a small variation of the sweep 
parameters. Ref.~\cite{fg2} calculated the qubit transition amplitude to all
orders in the non-adiabatic coupling. The result found there can be 
re-expressed as the following diagrammatic series: 
\begin{equation} 
\setlength{\unitlength}{0.05in}
T_{-}(t) \hspace{0.1in} = \hspace{0.1in}
                 \begin{picture}(10,4) 
                       \put(10,-1.5){\vector(-1,0){3.25}}
                       \put(5,-1.5){\line(1,0){1.75}} 
                       \put(5,-1.5){\vector(0,1){3.25}}
                       \put(5,1.75){\line(0,1){1.75}} 
                       \put(5,3.5){\vector(-1,0){3.25}}
                       \put(0,3.5){\line(1,0){1.75}} 
                 \end{picture} 
\hspace{0.05in} +
      \begin{picture}(20,4) 
           \put(20,-1.5){\vector(-1,0){3.25}}
           \put(15,-1.5){\line(1,0){1.75}} 
           \put(15,-1.5){\vector(0,1){3.25}}
           \put(15,1.75){\line(0,1){1.75}} 
           \put(15,3.5){\vector(-1,0){3.25}}
           \put(10,3.5){\line(1,0){1.75}} 
           \put(10,3.5){\vector(0,-1){3.25}}
           \put(10,-1.5){\line(0,1){1.75}} 
           \put(10,-1.5){\vector(-1,0){3.25}}
           \put(5,-1.5){\line(1,0){1.75}} 
           \put(5,-1.5){\vector(0,1){3.25}}
           \put(5,1.75){\line(0,1){1.75}} 
           \put(5,3.5){\vector(-1,0){3.25}}
           \put(0,3.5){\line(1,0){1.75}} 
     \end{picture} 
\hspace{0.05in} +
           \begin{picture}(30,5)
              \put(30,-1.5){\vector(-1,0){3.25}}
              \put(25,-1.5){\line(1,0){1.75}}
              \put(25,-1.5){\vector(0,1){3.25}}
              \put(25,1.75){\line(0,1){1.75}}
              \put(25,3.5){\vector(-1,0){3.25}}
              \put(20,3.5){\line(1,0){1.75}}
              \put(20,-1.5){\vector(-1,0){3.25}}
              \put(20,3.5){\vector(0,-1){3.25}}
              \put(20,-1.5){\line(0,1){1.75}}
              \put(15,-1.5){\line(1,0){1.75}}
              \put(15,-1.5){\vector(0,1){3.25}}
              \put(15,1.75){\line(0,1){1.75}}
              \put(15,3.5){\vector(-1,0){3.25}}
              \put(10,3.5){\line(1,0){1.75}}
              \put(10,3.5){\vector(0,-1){3.25}}
              \put(10,-1.5){\line(0,1){1.75}}
              \put(10,-1.5){\vector(-1,0){3.25}}
              \put(5,-1.5){\line(1,0){1.75}}
              \put(5,-1.5){\vector(0,1){3.25}}
              \put(5,1.75){\line(0,1){1.75}}
              \put(5,3.5){\vector(-1,0){3.25}}
              \put(0,3.5){\line(1,0){1.75}}
           \end{picture}
\hspace{0.1in} + \hspace{0.1in} \cdots \hspace{0.1in} . 
\label{diagser} 
\end{equation} 
Lower (upper) lines correspond to propagation in the negative (positive) 
energy-level, and the vertical lines correspond to transitions between the 
two energy-levels. The calculation sums the probability amplitudes for all
interfering alternatives \cite{f&h} that allow the qubit to end up in the 
positive energy-level given that it was initially in the negative energy-level.
As we have seen, varying the TRP sweep parameters varies the time 
separating the resonances. This in turn changes the value of each diagram 
in Eq.~(\ref{diagser}), and thus alters the interference between the 
alternative transition pathways. It is the sensitivity of the individual
alternatives/diagrams to the time separation of the resonances that 
allows TRP to manipulate this quantum interference. Zwanziger et al.\ 
\cite{zwan2} observed these interference effects in the transition 
probability using NMR and found quantitative agreement between theory 
and experiment. It is this link between interfering quantum alternatives 
and the TRP sweep parameters that we believe underlies the ability of 
TRP to drive high-fidelity non-adiabatic one- and two-qubit gates.

\subsection{Universal Quantum Gate Set}
\label{sec2.2}

The universal set of quantum gates $\calGU$ that is of interest here 
consists of the one-qubit Hadamard and NOT gates, together with 
variants of the one-qubit $\pi /8$ and phase gates, and the two-qubit
controlled-phase gate. Operator expressions for these gates are:
(1)~Hadamard: $U_{h}=(1/\sqrt{2})\left(\, \sigma_{z}+
     \sigma_{x}\right)$; (2)~NOT: $U_{not} = \sigma_{x}$;
(3)~Modified $\pi /8$: $V_{\pi /8} = \cos\left( \pi /8\right)\,
\sigma_{x} -\sin\left(\pi /8\right)\,\sigma_{y}$;
(4)~Modified phase: $V_{p} = (1/\sqrt{2})\left(\, \sigma_{x}
 -\sigma_{y}\,\right)$; and (5)~Modified controlled-phase:
$V_{cp} = (1/2)\left[ \left( I^{1}+\sigma_{z}^{1}\right) I^{2}
  - \left(I^{1}-\sigma_{z}^{1}\right)\sigma_{z}^{2}\right]$.
The universality of $\calGU$ was demonstrated in Ref.~\cite{lhg2}
by showing that its gates could construct the well-known universal 
set comprised of the Hadamard, phase, $\pi /8$, and CNOT gates.

\subsection{Simulation Procedure} 
\label{sec2.3} 
\noindent As is well-known, the Schrodinger dynamics is driven by a
Hamiltonian $H(t)$ that causes a unitary transformation $U$ to be 
applied to an initial quantum state $|\psi\rangle$. In this paper, it is 
assumed that the Hamiltonian $H(t)$ contains terms that Zeeman-couple 
each qubit to the TRP control-field $\bfF (t)$. Assigning values to the 
TRP sweep parameters $(a,b,B,T_{0})$ fixes the control-field $\bfF (t)$, 
and in turn, the actual unitary transformation $U_{a}$ applied to 
$|\psi\rangle$. The task  is to find TRP sweep parameter values that
produce an applied gate $U_{a}$ that approximates a desired target 
gate $U_{t}$ sufficiently closely that its error probability (defined below)
ideally satisfies $P_{e}<10^{-4}$. In the following, the target gate $U_{t}$ 
will be one of the gates in the universal set $\calGU$. Since $\calGU$ 
contains only one- and two-qubit gates, our simulations will only involve 
one- and two-qubit systems.

For the \textit{one-qubit simulations}, the Hamiltonian $H_{1}(t)$ is the 
Zeeman Hamiltonian $H_{z}(t)$ introduced in Section~\ref{sec2.1}. 
Ref.~\cite{lhg1} showed that it can be written in the following 
dimensionless form:
\begin{equation}
\calH_{1} (\tau ) = (1/\lambda)\,\left\{ -\tau\sigma_{z} 
                                 -\cos\phi_{4}(\tau )\sigma_{x} 
                                   -\sin\phi_{4}(\tau )\sigma_{y}\right\} .
\label{oneqbtHam}
\end{equation}
Here: $\tau = (a/b)t$; $\lambda = \hbar a/b^{2}$; and for quartic twist,
$\phi_{4}(\tau ) = (\eta_{4}/2\lambda )\tau^{4}$, with $\eta_{4}=\hbar 
Bb^{2}/a^{3}$.

For the \textit{two-qubit simulations}, the Hamiltonian $H_{2}(t)$ 
contains terms that Zeeman-couple each qubit to the TRP control-field, 
and an Ising interaction term that couples the two qubits. Alternative 
two-qubit interactions can easily be considered, though all simulation 
results presented below assume an Ising interaction between the qubits. 
To break a resonance-frequency degeneracy $\omega_{12}=\omega_{34}$
for transitions between, respectively, the ground and first-excited states 
($E_{1}\leftrightarrow E_{2}$) and the second- and third excited states 
($E_{3}\leftrightarrow E_{4}$), the term $c_{4}|E_{4}(t )\rangle\langle 
E_{4}(t )|$ was added to $H_{2}(t)$. Combining all of these remarks, we 
arrive at the following (dimensionless) two-qubit Hamiltonian \cite{lhg2}:
\begin{eqnarray}
\calH_{2}(\tau ) & = & 
               \left[ -(d_{1}+d_{2})/2+\tau/\lambda\right]\sigma_{z}^{1} 
                  -(d_{3}/\lambda )\left[\cos\phi_{4}\sigma_{x}^{1} +
                      \sin\phi_{4}\sigma_{y}^{1}\right]\nonumber\\
   & &     \hspace{0.275in}       +\left[ -d_{2}/2+\tau/\lambda\right]
                   \sigma_{z}^{2} 
             \hspace{0.075in}  -  (1/\lambda )\left[\cos\phi_{4}\sigma_{x}^{2} 
                   +       \sin\phi_{4}\sigma_{y}^{2}\right]\nonumber\\
   & &     \hspace{0.55in}        -(\pi d_{4}/2)\sigma_{z}^{1}\sigma_{z}^{2} 
                  \hspace{0.1in} +c_{4}|E_{4}(\tau )\rangle\langle 
                        E_{4}(\tau )| .
\label{twoqbtHam}
\end{eqnarray}
Here: (i)~$b_{i} = \hbar\gamma_{i}B_{rf}/2$, $\omega_{i}=\gamma_{i}
B_{0}$, $\gamma_{i}$ is the coupling constant for qubit $i$, and $i=1,2$; 
(ii)~$\tau = (a/b_{2})t$, $\lambda = \hbar a/b_{2}^{2}$, and $\eta_{4}=
\hbar Bb_{2}^{2}/a^{3}$; and (iii)~$d_{1}=(\omega_{1}-\omega_{2})
b_{2}/a$, $d_{2}=(\Delta /a)b_{2}$, $d_{3} = b_{1}/b_{2}$, and 
$d_{4}=(J/a)b_{2}$, where $\Delta$ is a detuning parameter \cite{lhg2}.

The numerical simulations assign values to the TRP sweep parameters and 
then integrate the Schrodinger equation to obtain the unitary transformation
$U_{a}$ produced by the resulting TRP sweep. Given $U_{a}$, $U_{t}$, and the 
initial state $|\psi\rangle$, it is possible to work out \cite{lhg1} the error 
probability $P_{e}(\psi )$ for the TRP final state $|\psi_{a}\rangle =
U_{a}|\psi\rangle$, relative to the target final state $|\psi_{t}\rangle 
= U_{t}|\psi\rangle$. The gate error probability $P_{e}$ is defined to be 
the worst-case value of $P_{e}(\psi )$: $P_{e}\equiv \max_{|\psi\rangle}
P_{e}(\psi )$. Introducing the positive operator $P = 
\left( U_{a}^{\dagger}-U_{t}^{\dagger}\right)\left( U_{a}-U_{t}
        \right) $,
Ref.~\cite{lhg1} showed that the error probability $P_{e}$ satisfies the 
upper bound $P_{e}\leq Tr\, P$. Once $U_{a}$ 
is known, $Tr\, P$ is easily evaluated, and so it is a convenient proxy 
for $P_{e}$ which is harder to calculate. $Tr\, P$ also has the virtue of 
being directly related to the gate fidelity $\calF_{n} = \left(1/2^{n}\right)
\, Re\left[\, Tr\left( U_{a}^{\dagger}U_{t}\right)\,\right]$ ,
where $n$ is the number of qubits acted on by the gate. It is 
straightforward to show \cite{lhg2} that $\calF_{n} = 1 -
\left(1/2^{n+1}\right)\, Tr\, P$.
The simulations calculate $Tr\, P$, which is then used to upper bound 
the gate error probability $P_{e}$. Note that 
minimizing $Tr\, P$ is equivalent to maximizing the gate fidelity $\calF$.

\section{Optimizing Gate Fidelity and Robustness}
\label{sec3}

\noindent To find TRP sweep parameter values that yield highly accurate 
non-adiabatic quantum gates, it proved necessary to combine the numerical 
simulations with function minimization algorithms \cite{numrec} that search 
for sweep parameter values that minimize the $Tr\, P$ upper bound. The 
multi-dimensional downhill simplex method was used for the one-qubit gates, 
while simulated annealing was used for the two-qubit modified controlled-phase 
gate. This optimization procedure produced one-qubit gates that operate with 
error probabilities satisfying $P_{e}<10^{-4}$ \cite{lhg1}. However, for the 
two-qubit modified controlled-phase gate
$V_{cp}$, simulated annealing was only able to find sweep parameter values 
that gave $P_{e}\leq 1.27\times 10^{-3}$ \cite{lhg2}. To further improve
the performance of this two-qubit gate, Ref.~\cite{lg1} incorporated
the group-symmetrized evolution of Ref.~\cite{zan} to obtain a modified
controlled-phase gate with $P_{e}< 10^{-4}$. Group-symmetrized evolution 
is a form of dynamical decoupling that produces an effective dynamics that is 
invariant under the action of a finite symmetry group $\calG$. Ref.~\cite{lg1}
identified the finite group~$\calG$ with the symmetry group of $V_{cp}$, and
then used the procedure of Ref.~\cite{zan} to filter out the 
$\mathcal{G}$-noninvariant part of the TRP dynamics. As the 
$\mathcal{G}$-noninvariant dynamics is manifestly bad dynamics relative to
$V_{cp}$, group-symmetrized TRP yields a better approximation to $V_{cp}$,
and produces a smaller gate error probability. The reader is directed to 
Ref.~\cite{lg1} for an detailed explanation of how group-symmetrized evolution 
is incorporated into a TRP sweep. As noted in Section~\ref{sec1}, although this
combination of simulation, optimization, and group-symmetrized evolution
yielded a universal set of quantum gates that operate with the anticipated 
accuracy needed for fault-tolerant quantum computing, it was found that the 
one-qubit gates in $\calGU$ could only achieve this level of accuracy if the 
TRP sweep parameters were controllable to $17$-bit ($1$ part in 
$10^{5}$) precision. Such precision is beyond the capabilities of present-day 
commercially available arbitrary waveform generators. Thus, if TRP is to 
provide a viable approach to high-fidelity universal quantum control, a way
must be found to improve the robustness of the TRP-generated one-qubit
gates.  In this Section we show how to modify our optimization procedure so 
that it rewards robust gate operation, while still minimizing (maximizing) $Tr\, P$
(gate fidelity).

The optimization procedure just described searches for parameter 
values $x=(x_{1},\ldots , x_{n})$ that minimize the cost function
\begin{equation}
\kappa (x) = Tr\, P(x_{1},\ldots ,x_{n}) .
\label{costfcn}
\end{equation}
As described in Section~\ref{sec2.3}, for the one-qubit gates in $\calGU$, 
$n=2$, and $x_{1}=\lambda$ and $x_{2}=\eta_{4}$. For $V_{cp}$, $Tr\, P$ 
depends on seven parameters $(\lambda ,\eta_{4},c_{4},d_{1},\ldots ,d_{4})$. 
However, for group-symmetrized TRP, only $c_{4}$ and $d_{4}$ are critical 
parameters \cite{lg1} and so, effectively, $n=2$ and $(x_{1}=c_{4} , x_{2}
=d_{4})$. Thus $n=2$ in the cases of interest, although we will consider 
arbitrary $n$ in the following analysis.

Let $\overline{x} =(\xbar_{1},\ldots ,\xbar_{n})$ and $\kbar = 
\kappa (\xbar )$ denote the parameter and cost function values, respectively,
at an optimization minimum. Suppose we vary the parameters slightly
away from $\xbar$: $x_{i} = \xbar_{i} + \delta x_{i} \:(i=1,\ldots , n)$.
Taylor-series expanding the cost function $\kappa (x)$ about $\xbar$ gives
\begin{equation}
\kappa (\xbar + \delta x) = \kbar +\delta^{2}\kappa + \mathcal{O}
                                               (\delta^{3}x),
\label{taylrcost}
\end{equation}
where the first-order variation vanishes since $\xbar$ specifies a minimum of
$\kappa (x)$.  In Eq.~(\ref{taylrcost}),
\begin{equation}
\delta^{2}\kappa = \frac{1}{2}\sum_{i,j=1}^{n}\delta\xi_{i}H_{ij}
                                    \delta\xi_{j} ,
\label{secndvar}
\end{equation}
where $\xi_{i} = x_{i}/\xbar_{i}$, and $H$ is the Hessian of the cost 
function $\kappa$ whose matrix elements are:
\begin{equation}
H_{ij} = \left. \frac{\partial^{2}\kappa}{\partial\xi_{i}\partial\xi_{j}}
                              \right|_{x=\xbar} .
\label{Hessdef}
\end{equation}
It follows from Eq.~(\ref{taylrcost}) that the Hessian $H$ determines how
rapidly the cost function $\kappa = Tr\, P$ varies in the vicinity of the
minimum $\xbar$. It will thus play an central role in our robustness analysis.

Examination of the Tables in Ref.~\cite{lhg2} shows that $\delta^{2}\kappa
\sim (30-100)\,\kbar$ when $\xbar_{i}$ is varied in its fifth significant figure.
As noted earlier, more robust gate performance is desired. 
Experimental control of the TRP sweep parameters to four significant figures 
($14$-bit precision) is possible and so we can consider a gate to have robust 
performance if $\delta^{2}\kappa\sim\kbar$ when $\xbar_{i}$ is varied in its 
fourth significant figure. This condition can be transformed into a condition 
on the $l_{1}$-norm \cite{hrn&jhn} of the Hessian $H$: 
$||\, H\, ||_{1} = \sum_{i,j=1}^{n} |H_{ij}|$ .
To see this, suppose that to four significant figure precision, $\xbar_{i}= 
\xbar_{i}^{0}.\xbar_{i}^{1}\xbar_{i}^{2}\xbar_{i}^{3}\times 10^{e_{i}}$. 
Varying $\xbar_{i}$ in its fourth significant figure $(\xbar_{i}\rightarrow\xbar_{i}+
0.001\times 10^{e_{i}})$ gives
\begin{equation}
\delta\xi_{i} =\frac{\delta x_{i}}{\xbar_{i}} = 
\frac{0.001}{\xbar_{i}^{0}.\xbar_{i}^{1}\xbar_{i}^{2}\xbar_{i}^{3}} .
\label{delxi}
\end{equation}
Combining $\delta^{2}\kappa\sim\kbar$ with Eqs.~(\ref{secndvar}) and 
(\ref{delxi}) gives
\begin{equation}
\kbar = \left( 5\times 10^{-7}\right)\sum_{i,j=1}^{n}
     \frac{H_{ij}}{(\xbar_{i}^{0}.\xbar_{i}^{1}\xbar_{i}^{2}\xbar_{i}^{3})
      (\xbar_{j}^{0}.\xbar_{j}^{1}\xbar_{j}^{2}\xbar_{i}^{3})} .
\label{intres}
\end{equation}
To arrive at a representative robustness condition, note that for the gate 
simulations presented in Ref.~\cite{lg1}, $\kbar\sim 5\times 10^{-5}$, 
and $\xbar_{i}^{0}.\xbar_{i}^{1}\xbar_{i}^{2}\xbar_{i}^{3}\sim 5$ for 
all $i$. For such representative values, Eq.~(\ref{intres}) gives 
$\sum_{i,j=1}^{n} H_{ij} = 2500$. Noting that $||\, H\, ||_{1}\geq 
\sum_{i,j=1}^{n}H_{ij}$, this gives $||\, H\, ||_{1}\geq 2500$. Thus robust 
gate performance will be obtained if the $l_{1}$-norm of the Hessian 
$H$ satisfies $||\, H\, ||_{1}\sim 2500$. This condition allows us to introduce 
a penalty function $\calP (x)$ for gate robustness, where $\calP (x) = 
\left( ||\, H\, ||_{1} - 2500\right)^{2}$ when $||\, H\, ||_{1}\ge 2500$; and
is zero otherwise. The penalty function $\calP (x)$ is small when a gate is 
operating robustly, and increases as gate performance becomes progressively 
less robust. To sensitize our optimization procedure to gate robustness, we add 
$\calP (x)$ to the cost function $\kappa (x)$:
\begin{equation}
\kappa (x) = Tr\, P(x) + r\,\calP (x).
\label{newcostfcn}
\end{equation}
Here $r$ is a parameter that specifies how heavily gate robustness is
weighted during the sweep parameter optimization. When $r=0$, $\kappa (x)$
reverts to our previous cost function (Eq.~(\ref{costfcn})), and to 
a robustness-insensitive optimization. In the following Section we use the 
new cost function (Eq.~(\ref{newcostfcn})) to harden the robustness of the 
one-qubit gates in the universal set $\calGU$.

\section{Gate Results} 
\label{sec4} 
\underline{\textit{One-Qubit Gates:}} Here we present our simulation 
results for the one-qubit gates in $\calGU$, with
the sweep parameter optimization based on the new cost function
$\kappa (x)$ appearing in Eq.~(\ref{newcostfcn}). Table~\ref{table1}
\begin{table}
\tbl{Simulation results for the one-qubit gates in $\calGU$.
The error probability for each gate satisfies $P_{e}\leq Tr\, P$.}
{\begin{tabular}{|c|c|c|c|c|}
\hline
Gate\rule[0.5ex]{0cm}{2ex} & $\lambda$ & $\eta_{4}$ & $Tr\, P$ &
  $\calF$\\
\hline
NOT\rule[0.5ex]{0cm}{2ex} & $6.965$ & $2.189\times 10^{-4}$ & 
  $6.27\times 10^{-5}$ & $0.9999\, 8$\\
Hadamard & $7.820$ & $1.792\times 10^{-4}$ & $1.12\times 10^{-4}$ 
   & $0.9999\, 7$\\
Modified $\pi /8$ & $8.465$ & $1.675\times 10^{-4}$ & $2.13\times 10^{-4}$ &
  $0.9999\, 5$\\
Modified phase & $8.073$ & $1.666\times 10^{-4}$ & $4.62\times 10^{-4}$ 
  & $0.9998\, 8$\\
\hline
\end{tabular}}
\label{table1}
\end{table}
gives the optimum values for the dimensionless sweep parameters $\lambda$ and
$\eta_{4}$ for each of the one-qubit gates in $\calGU$. The connection between 
the dimensionless and dimensionful sweep parameters appears below 
Eq.~(\ref{oneqbtHam}). Note that all one-qubit simulations were done with 
dimensionless inversion time $\tau_{0}=80.000$. Table~\ref{table1} also gives 
the $Tr\, P$ upper bound on the gate error probability $P_{e}\leq Tr\, P$ and 
the gate fidelity $\calF$. We see that all one-qubit gates operate with very 
high-fidelity, with two gates operating with gate error probabilities satisfying 
$P_{e}\leq 1.12\times 10^{-4}$, and the remaining two gates satisfying $2.13
\times 10^{-4}\leq P_{e}\leq 4.62\times 10^{-4}$. Thus all one-qubit gates in 
$\calGU$ operate at, or nearly at, the anticipated accuracy needed for 
fault-tolerant quantum computing. We now show that, due to our modified 
optimization procedure, the one-qubit gates are able to achieve the performance
given in Table~\ref{table1} if the TRP sweep parameters are controllable to 
$14$-bit ($1$~part in $10^{4}$) precision. Table~\ref{table2}
\begin{table}
\tbl{Sensitivity of $Tr\, P$ to small variation of $\lambda$
and $\eta_{4}$ for the one-qubit Hadamard gate. The three 
left-most (right-most) columns correspond to varying $\lambda$ ($\eta_{4}$) in
its fourth significant digit at fixed $\eta_{4}$ ($\lambda$).}
{\begin{tabular}{|ccc||ccc|}
\hline
$\eta_{4}$ & $\lambda$\rule[0.5ex]{0cm}{2ex} & $Tr\, P$ & $\lambda$ &
$\eta_{4}$ & $Tr\, P$ \\
\hline
$1.792\times 10^{-4}$\rule[0.5ex]{0cm}{2ex} & $7.819$ & $8.05\times 10^{-4}$ 
  & $7.820$ & $1.791\times 10^{-4}$ & $2.86\times 10^{-2}$\\
$1.792\times 10^{-4}$ & $7.820$ & $1.12\times 10^{-4}$ & $7.820$ &
 $1.792\times 10^{-4}$ & $1.12\times 10^{-4}$\\
$1.792\times 10^{-4}$ & $7.821$ & $2.07\times 10^{-3}$ & $7.820$ &
 $1.793\times 10^{-4}$ & $3.11\times 10^{-2}$\\
\hline
\end{tabular}}
\label{table2}
\end{table}
shows how $Tr\, P$ varies for the Hadamard gate as we vary either 
$\lambda$ or $\eta_{4}$ in its fourth significant digit. Similar behavior 
occurs with the other one-qubit  gates in $\calGU$ and so we do not display 
corresponding Tables for these gates. Note that when a hardware 
parameter $p$ is said to have $14$-bit precision, it means that it is 
specifiable to $4$ significant figures. Specifically, if one wants 
$\eta_{4}=1.792\times 10^{-4}$, the hardware gives that value and \textit{not\/} 
$1.791\times 10^{-4}$ or $1.793\times 10^{-4}$. Thus with $14$-bit precision
sweep parameters one can hit the optimum sweep parameter values and achieve 
the performance shown in Table~\ref{table1}. However, with \textit{less\/} than 
$14$-bit precision, gate performance will be washed out over the entries in the 
$Tr\, P$ columns of Table~\ref{table2}, and gate performance will not reach the 
level of Table~\ref{table1}. This is to be compared with Ref.~\cite{lg1} where 
gates with $P_{e}< 10^{-4}$ required sweep parameters with $17$-bit 
precision. The price paid for this enhanced robustness is a slight loss in gate 
fidelity compared to the one-qubit gates presented in Ref.~\cite{lg1}.

\underline{\textit{Modified Controlled-Phase Gate:}} As shown in Ref.~\cite{lg1}, 
group-symmetrized TRP is able to
produce a two-qubit modified controlled-phase gate $V_{cp}$ which has
$Tr\, P = 8.87\times 10^{-5}$, corresponding gate fidelity $\calF = 
0.9999\, 9$, and error probability satisfying
$P_{e}\leq 8.87\times 10^{-5}$. Ref.~\cite{lg1} also showed that this
level of accuracy could be achieved with control parameters specified with
$14$-bit ($1$ part in $10^{4}$) precision. Thus $V_{cp}$ already operates
at the anticipated accuracy needed for fault-tolerant quantum computing
with precision demands that are within reach of currently available arbitrary
waveform generators. There is thus no need to re-do the optimization of 
this gate since it is already both sufficiently accurate and robust. The sweep
parameter values found in Ref.~\cite{lg1} that produce this gate are included
here for completeness: $\lambda = 5.04$, $\eta_{4}=3.0\times 10^{-4}$,
$\tau_{0} = 120.00$, $c_{4} = 2.173$, $d_{1} = 99.3$, $d_{2}= 0.0$, 
$d_{3}=-0.41$, and $d_{4} = 0.8347$.

\section{Discussion}
\label{sec5}
\noindent We have presented a general approach for enhancing
the robustness of an arbitrary quantum gate and have applied this approach 
to the one-qubit gates implemented using twisted rapid passage (TRP). We have 
shown that the resulting gates operate with error probabilities (fidelities) in 
the range $6.27\times 10^{-5} < P_{e} < 4.62\times 10^{-4}$ 
($0.9998\, 8 < \calF < 0.9999\, 9$), while only requiring TRP sweep parameters 
that are controllable to $14$-bit ($1$ part in $10^{4}$) precision. In 
conjunction with the group-symmetrized two-qubit gate presented in 
Ref.~\cite{lg1}, our robustness enhancing procedure has 
yielded a universal set of quantum gates that: (i)~operate with fidelities that 
are at, or nearly at, the anticipated accuracies needed for fault-tolerant 
quantum computing; and (ii)~are sufficiently robust to be within reach of 
commercially available arbitrary waveform generators (AWG) \cite{tek}.
These results suggest the feasibility of an experimental study (see below) 
of TRP-based high-fidelity universal quantum control. 

In Ref.~\cite{lg1} we showed how TRP could be used to
produce a universal set of quantum gates that operate with error probabilities
(fidelities) satisfying $P_{e}<10^{-4}$ ($\calF > 0.9999$). However, it was 
found that the one-qubit gates in this universal set required the TRP sweep 
parameters to be controllable to $17$-bit ($1$~part in $10^{5}$) precision 
which is beyond the reach of commercially available AWG. Using the new 
optimization procedure presented in Section~\ref{sec3}, we have been 
able to increase the robustness of the one-qubit TRP gates to $14$-bit 
precision which, as noted above, is within reach of commercially available 
AWG \cite{tek}. The cost of this improvement in gate robustness, however, 
was a slight reduction in the fidelity of the one-qubit TRP gates.

It is worth noting that earlier work \cite{fg1,lg1,lhg2,lhg1} showed how 
TRP sweeps could be applied to NMR, atomic, and superconducting qubits, as 
well as to spin-based qubits in quantum dots; while Ref.~\cite{lhg1} described 
how quantum state tomography could be used to test the performance of the 
TRP-generated universal quantum gate set. The reader is directed to those 
papers for further discussion.

A number of directions for future work are possible. (1)~Possibly
the most important at this time is finding a way to improve the accuracy of 
the TRP one-qubit gates so that \textit{all\/} have error probabilities 
satisfying $P_{e}< 10^{-4}$, while still requiring no more than $14$-bit 
precision on the TRP sweep parameters. Work on this is underway. (2)~In 
previous work we have studied a number of forms of polynomial, as well as 
periodic, twist \cite{lgm}. To date, we have found that quartic twist provides 
best all-around performance when it comes to making the gates in $\calGU$. 
We do not at present have arguments to explain why this is so. We have 
developed a framework for studying the optimal form of the TRP twist profile 
$\phi (t)$ based on quantum optimal control theory. We plan to examine this 
important question in future work. (3)~Finally, it would be interesting to study 
the impact of using a non-Ising two-qubit interaction on the performance of 
the TRP two-qubit gate. 

\section*{Acknowledgements}
This research was supported in part by the National Science Foundation through
TeraGrid computational resources provided by NCAR under grant TG-PHY100038.
One of us (F.G.) thanks T. Howell III for continued support.

\markboth{Ran Li and Frank Gaitan}{Journal of Modern Optics}
\end{document}